\documentclass{aastex63}

\usepackage{amsmath,amssymb,bbold,mathrsfs,bm}
\usepackage{textcase}
\usepackage{subfigure}

\usepackage[nonumberlist,nosuper]{glossaries}
\setacronymstyle{long-short}

\newacronym{rt}{RT}{radiation transfer}
\newacronym{rte}{RTE}{radiation transfer equations}
\newacronym{see}{SEE}{statistical equilibrium equations}
\newacronym{se}{SE}{statistical equilibrium}
\newacronym{los}{LOS}{line of sight}
\newacronym{1d}{1D}{one-dimensional}
\newacronym{3d}{3D}{three-dimensional}
\newacronym{prd}{PRD}{partial frequency redistribution}
\newacronym{crd}{CRD}{complete frequency redistribution}
\newacronym{mo}{MO}{magneto-optical}
\newacronym{ali}{ALI}{accelerated $\Lambda$-iteration}

\newcommand{\TAUE}[1]{${\rm log_{10}}(\tau_{500})=#1$}
\newcommand{\TAUA}[1]{${\rm log_{10}}(\tau_{500})\approx#1$}
\newcommand{\SII}{${\rm [pJ\ m^{-2}\ s^{-1}\ Str^{-1}\ Hz^{-1}]}$}

\begin{document}

\title{TIC: A Stokes inversion code for scattering polarization \\
       with partial frequency redistribution and arbitrary magnetic fields}
\shorttitle{TIC: Stokes inversion with PRD and Scattering}

\author{H.\ Li}
\affil{Instituto de Astrof\'{\i}sica de Canarias, E-38205 La Laguna,
       Tenerife, Spain}
\affil{Departamento de Astrof\'\i sica, Universidad de La Laguna, E-38206
       La Laguna, Tenerife, Spain}
       
\author{T.\ del Pino Alem\'an}
\affil{Instituto de Astrof\'{\i}sica de Canarias, E-38205 La Laguna,
       Tenerife, Spain}
\affil{Departamento de Astrof\'\i sica, Universidad de La Laguna, E-38206
       La Laguna, Tenerife, Spain}
       
\author{J.\ Trujillo Bueno}
\altaffiliation{Affiliate Scientist of the National Center for Atmospheric Research, Boulder, CO, USA.}
\affil{Instituto de Astrof\'{\i}sica de Canarias, E-38205 La Laguna,
       Tenerife, Spain}
\affil{Departamento de Astrof\'\i sica, Universidad de La Laguna, E-38206
       La Laguna, Tenerife, Spain}
\affil{Consejo Superior de Investigaciones Cient\'{\i}ficas, Spain}

\author{R. Casini}
\affil{High Altitude Observatory, National Center for Atmospheric
 Research.\break P.O.~Box 3000, Boulder, CO 80307-3000, U.S.A.\break}

\begin{abstract}
We present the Tenerife Inversion Code (TIC), which has been developed
to infer the magnetic and plasma properties of the solar chromosphere and transition   
region via full-Stokes inversion of polarized spectral lines. The code is based  
on the HanleRT forward engine, which takes into account many of the physical mechanisms  
that are critical for a proper modeling of the Stokes profiles of spectral lines
originating in the tenuous and highly dynamic plasmas of the chromosphere and
transition region: quantum level population imbalance and interference (atomic polarization), 
frequency coherence effects in polarized resonance scattering (partial frequency redistribution), 
and the impact of arbitrary magnetic fields on the atomic polarization and the radiation field. 
We present first results of atmospheric and magnetic inversions, 
and discuss future developments for the project.
\end{abstract}


\section{Introduction} \label{sec:intro}
The solar chromosphere lies between 
the relatively cold, few thousand kelvin photosphere
and the hot, million kelvin corona. This extended region spans about nine pressure
scale heights and, even though it shows lower temperatures than the overlying corona,
its larger density requires a comparatively larger energy deposit for its maintenance 
\cite[e.g.,][]{Carlsson2019ARA&A}. The magnetic fields that permeate the
solar atmosphere and dominate the structuring of the low-$\beta$ plasma
(where $\beta$ is the ratio of gas to magnetic pressure) are key to understanding the
coupling of its various layers and how the energy that is produced in the inner
layers of the solar atmosphere is transported outward, and converted into
heating of the chromosphere and corona. 
One of the main challenges of solar physics faced nowadays is the
determination of the magnetic field in the upper solar atmosphere \citep[e.g., the review 
by][]{Trujillo2022}. 
The polarization of the electromagnetic radiation emerging from the solar atmosphere carries information
on the physical properties of the emitting plasma, including the magnetic field.
Therefore, the study of the polarized solar spectrum is critical to uncovering the
properties of the magnetic field, which in turn allows us to better understand the
physical processes taking place in the solar atmosphere.

Stokes inversion techniques are a necessary tool to infer the physical
properties of the solar plasma from spectropolarimetric observations. During
the last decades, sophisticated Stokes inversion codes have been developed and
applied to solar observations
\cite[see the reviews by][and references therein]{Iniesta2016LRSP,Lagg2017SSRv, delaCruz2017SSRv},
but there is still need for progress especially regarding the interpretation 
of spectropolarimetric profiles of chromospheric and coronal lines.
As the plasma density decreases with height in the solar atmosphere, collisional
processes become less and less important in determining the excitation and ionization
balance of atoms, and spectral lines form outside of thermodynamical equilibrium (non-LTE).
Moreover, photon coherence in scattering processes becomes increasingly important
(the so called partial frequency redistribution, PRD), dramatically impacting
polarized spectral line formation.
While some of today's Stokes inversion 
codes can account for non-LTE effects (e.g.,
NICOLE, \citealt{Socas-Navarro2000ApJ,Socas-Navarro2015A&A}, 
and SNAPI, \citealt{Milic2018A&A}) and even
\gls*{prd} effects (STiC, \citealt{delaCruz2019A&A}, and DeSIRe, \citealt{DeSIRe}), 
these codes only account for Zeeman-effect polarization.

In the chromosphere, however, the excitation state of the atoms is affected by
the anisotropy of the incident radiation and the atomic levels become polarized (through
population imbalance and quantum coherence among the magnetic sublevels in the atom).
This condition produces the so-called scattering polarization of the re-emitted
radiation, which is sensitive to the magnetic field via the Hanle effect
\cite[e.g.,][]{TrujilloBueno2001,LL04}. 
There are both theoretical and numerical complexities implied by modeling atomic polarization,
e.g., the necessity to keep track of the geometry with respect to the propagation
directions of radiation within the atmosphere,
instead of just the current \gls*{los}, and to account
for the breaking of axial symmetry in the pumping radiation field caused by horizontal
radiation transfer, which is also a source of scattering polarization
\citep[e.g.,][]{MansoTrujillo2011}.
A well-known inversion code that includes some  
of these physical ingredients
is HAZEL \citep{AsensioRamos2008ApJ}, although its present treatment 
of \gls*{rt} is
not applicable to the spectral lines originating in optically thick regions
of the solar chromosphere. Moreover, it does not account for \gls*{prd} effects.

Most of the inversion techniques accounting for \gls*{rt} and non-LTE rely on the
computation of response functions \citep{Magain1986A&A,LL04}. The response function 
can be computed numerically (e.g., NICOLE and STiC) or analytically
(e.g., the SNAPI code). The former method is very time consuming, and it typically
dominates the computing requirements of the inversion procedure.
While the latter is theoretically much faster, its speed and performance have only
been tested without atomic polarization.
Moreover, the analytical calculation of response functions requires to explicitly
account for all interdependencies in the \gls*{se} equations. This is already
a complex problem when only accounting for atomic populations \citep{Milic2017A&A}
and the complexity dramatically increases when accounting for atomic polarization.

Recently, the sounding rocket experiments CLASP and CLASP2
have provided unprecedented spectropolarimetric observations of the \ion{H}{1} Ly-$\alpha$
and the \ion{Mg}{2} h and k lines, respectively \citep{Kano+2017,Ishikawa2021}.
These observations have confirmed 
theoretical predictions based on the quantum theory of spectral line polarization
(\citealt{TrujilloBueno+2011,Belluzzi2012ApJ,Belluzzi2012ApJb,Stepan+2015,
AlsinaBallester2016ApJ,Tanausu2016ApJ,Mansoetal2019,Tanausu2020ApJ}),
showing clearly the impact of scattering polarization and the
\gls*{mo} and \gls*{prd} effects on the observed polarization profiles.
The same observations have also revealed some surprises, which  
have expanded our understanding of the scattering polarization in  
strong resonance lines \citep{Trujillo+2018}.  
It has thus become clear that there is a need for diagnostic tools that are capable
of taking into account all these necessary physical ingredients to model the
polarization of strong chromospheric spectral lines.

In this paper we present the Tenerife Inversion Code (TIC), which 
takes into account scattering polarization  
and the effects of \gls*{prd} and quantum level interference
in the presence of arbitrary magnetic fields
(from zero field to the complete Paschen-Back regime). To achieve this, TIC is based
on the HanleRT spectral synthesis
code \citep{Tanausu2016ApJ,Tanausu2020ApJ}. In \S~\ref{Scode} we describe the
inversion algorithm and its implementation. In \S~\ref{Stest} we show the
application of the TIC to spectral profiles obtained from the synthesis in
semi-empirical one-dimensional models in order to assess its performance in this necessary
but non-trivial test. We show applications both without and with added photon
noise. We finish this section with an application relying only on the
circular polarization profiles. Finally, we present our summary and
discussion in \S~\ref{Scon}, including a brief discussion on future developments
for the project.


\section{Inversion Code}\label{Scode}

We call inversion the semi-automatic or automatic process of inferring the
magnetized model atmosphere that, when input into its spectral synthesis module 
(hereafter, forward engine), gives the
best fit to a given observation. Of course, the inference is highly dependent
on the forward engine and how well it describes the physics of the generation
and transfer of polarized radiation. In this section we describe the main
characteristics of the TIC, a numerical code for the inversion of
Stokes profiles that takes into account the physical mechanisms necessary
for the modeling of the polarization in strong resonance chromospheric
spectral lines, namely, atomic polarization and \gls*{prd} effects in the
presence of magnetic fields of arbitrary strength.
TIC relies on 
HanleRT \citep{Tanausu2016ApJ,Tanausu2020ApJ} as its forward engine.


\subsection{Forward synthesis module}

HanleRT was developed by \citet{Tanausu2016ApJ, Tanausu2020ApJ} to solve
the problem of the generation and transfer of polarized radiation out of local
thermodynamical equilibrium (non-LTE) in one-dimensional plane-parallel atmospheric models,
taking into account the coherent scattering of radiation by polarized multi-term atoms
in arbitrary magnetic-fields \citep{Casini2014ApJ,Casini2017ApJ,Casini2017bApJ}.
Consequently, the TIC can take into account scattering polarization with \gls*{prd}
effects, and quantum level interference, for arbitrary magnetic fields.
Even though HanleRT implements the general angle-dependent
redistribution function to describe \gls*{prd} effects, in this paper we show
results obtained with the angle-averaged 
approximation \citep{Mihalas1970,Belluzzi2014ApJ} in
order to significantly reduce the total computational cost.


\subsection{Inversion module}\label{sec:inversionmod}

As any inversion code, the TIC finds the model atmosphere that produces the 
best fit to a given set of Stokes profiles via its forward engine.
To this end, the inversion code determines at each iteration step how the
physical parameters describing the current model atmosphere
must be modified in order to improve the fit. The magnitude of these
changes is determined from the response functions
\citep[e.g.,][]{Magain1986A&A,LL04,Uitenbroek2006ASPC} of the emergent Stokes profiles  
to perturbations in the physical parameters of the model 
in a set of chosen nodes sampling the atmosphere's height
stratification. This is the usual strategy employed by existing inversion codes.

The physical parameters of the model atmosphere used in the TIC are the plasma temperature ($T$), the
bulk velocity vector ($\bm{v}$), the micro-turbulent velocity ($v_{\rm turb}$),
the magnetic field vector ($\bm{B}$), and the gas pressure at the top boundary\footnote{
In order to derive the model's density stratification, we assume hydrostatic equilibrium
\citep[e.g.,][]{Mihalas1970}, reason why only the boundary value is needed once the temperature
stratification is known.} ($P$). In the spectral synthesis code, the number densities of the different atomic
species are computed by solving the equation of state with the method of \cite{Wittmann1974solphys}.

Usually, inversion codes retrieve the atmospheric stratification along the \gls*{los}
and, in particular, the bulk velocity along the same direction. However, when
accounting for scattering polarization, the orientation of the \gls*{los} with respect 
to the atmospheric structure becomes important. Therefore, 
even in a \gls*{1d} plane-parallel model atmosphere, we must distinguish between the vertical and LOS 
directions in the inversion.
Although we can include the three components of the macroscopic velocity vector, its horizontal 
component is a source of scattering polarization due to the breaking of axial symmetry 
\citep[e.g.,][]{Stepan2016ApJ,Tanausu2018ApJ,Jaume+2021}.
Because the inversions in this paper are performed using Stokes profiles calculated 
in known axially symmetric model atmospheres,
we assume that the horizontal component of the velocity field is zero,
preserving the axial symmetry, and we only invert its vertical component.

In order to find the best fit to a given set of Stokes profiles, we minimize a cost
function that accounts for both the difference between the data and the synthetic profiles
emerging from the inverted model, and additional constraints imposed on the
solution (regularizations). The cost function can thus be written as
\begin{equation}
	\chi^2=\frac{1}{4N_\lambda}\sum\limits_{i=1}^4\sum\limits_{j=1}^{N_\lambda}
	\left[\frac{I_i^{\rm obs}(\lambda_j)-I_i^{\rm syn}(\lambda_j)}{\sigma_{i}(\lambda_j)}\right]^2w_{i}^2
	+\sum\limits_{n=1}^N\alpha_nr^2_n(p) ,
\label{Eq1}
\end{equation}
where $N_\lambda$ is the number of wavelength points in the observed Stokes profiles.
$I_i^{\rm obs}(\lambda_j)$ and $I_i^{\rm syn}(\lambda_j)$
are the $i$-th Stokes parameters ($i = I$, $Q$, $U$, and $V$) at each $j$-th
wavelength position for the observed and synthetic profile, respectively.
$\sigma_{i}(\lambda_j)$ is the noise of each ``$i$'' Stokes parameter in the
observation, assumed to be Gaussian and, in general, dependent on the wavelength
index ``$j$''. $w_{i}$ is a weight for each Stokes parameter to account for the
expected difference in their order of magnitude for different physical
scenarios.\footnote{Note that the intensity in solar spectral lines is usually at
least two orders of magnitude larger than the
polarization signals and it would thus dominate the cost
function if no weights were added.} $r_n(p)$ is a regularization function weighted
by $\alpha_n$. 
These regularization functions include penalties on the first derivatives of the stratification 
of the model parameters (bulk velocity, micro-turbulent velocity, and magnetic field, 
favoring smooth stratifications over complex ones), 
penalties on the second derivatives of the stratification 
of the model parameters (temperature, favoring smooth gradients over complex ones), 
and penalties on deviations from a given value for a
model parameter (gas pressure at the top boundary, favoring solutions fulfilling 
some a priori knowledge). 
These regularization functions have been used in other inversion 
codes such as the STiC \citep{delaCruz2019A&A}.
The regularization function weights $\alpha_n$ are not kept constant during the inversion.
At the beginning of the inversion procedure, when $\chi^2$ can be several orders of
magnitude larger than the target value, we start with enhanced values of $\alpha_n$,
such that the regularization term in Eq.~\eqref{Eq1} is of the same order of magnitude
as the first term describing just the quality of the fit. As the inversion moves toward
smaller $\chi^2$, the regularization function weights are progressively reduced until they
reach the values that are specified for the inversion solution.

We have implemented the Levenberg-Marquardt algorithm \citep{NumericalRecipes} to
minimize the cost function in Eq.~\eqref{Eq1}, with the Hessian matrix approximated by
only using the first derivatives with respect to the model parameters (i.e. the
response functions). The Hessian being second order, we approximate the second
derivatives with the product of first derivatives.
We thus compute the response functions of all model parameters at
each iteration step. To that end, we perturb their node values and 
recompute the Stokes profiles to compare them with the ones resulting from the unperturbed 
model. We have implemented the calculation of these response functions with both 
central differences (compute the Stokes parameters perturbing both positively and 
negatively each parameter) and forward differences (compute the Stokes parameter 
perturbing each parameter only once). Although the former is more accurate, we have 
not found significant differences between the response functions retrieved by the 
two methods in our tests, and thus we apply the latter 
throughout this paper as it requires half the number of forward syntheses.

Regarding the damping parameter $\lambda$ of the Levenberg-Marquardt 
algorithm \citep[see][]{NumericalRecipes}, 
we have implemented a parabolic interpolation in order to optimize its value at each
iteration. This procedure, dubbed \emph{backtracking} method, is adopted also by
other inversion codes, such as the STiC and HAZEL.

In order to determine the correction to the model parameters we need to solve a
linear system of equations. We apply the modified singular value decomposition
(SVD) method proposed by \cite{RuizCobo1992ApJ}. We start the SVD with a relatively
small ($10^{-6}$) tolerance factor $\epsilon$ and check the resulting corrections.
If they are larger than a certain threshold (e.g., the correction to the
temperature is larger than $4\cdot10^3$~K or the correction to the velocity is
larger than $10$~km/s), we increase the tolerance factor and solve the system again,
repeating the procedure until the corrections comply with the pre-established
threshold or $\epsilon$ is of order $10^{-3}$.

The uncertainties in the inverted model parameters could also be estimated by applying
Bayes' theorem with a Monte Carlo method. However, the computational cost of each
forward solution makes this approach unfeasible. We thus estimate the uncertainties 
following the approach by \citet{SanchezAlmeida1997ApJ} \citep[see also][]{delToroIniesta2003} to
compute the variance of each model parameter at each node ($p$):
\begin{equation}
	\sigma_p^2 = \frac{2 \Delta\chi^2}{N_{\rm nodes}}H^{-1} ,
\label{Eq2}
\end{equation}
where $N_{\rm nodes}$ is the number of nodes, $\Delta\chi^2$ is the first term in the
right side of Eq.~\eqref{Eq1}, i.e., the $\chi^2$ without the regularization term,
and $H^{-1}$ is the inverse matrix of the Hessian matrix. $H^{-1}$ is simply approximated by
the inverse diagonal elements of $H$. 
Since these diagonal elements are in fact the squares of the response functions, the uncertainties
can thus be estimated directly from the values of the response functions, giving an
immediate estimation of how well the
inferred parameters are constrained. Even though this approach does not give accurate
uncertainties, it provides an idea of the sensitivity of the different physical parameters to
the changes in each of the nodes in the model atmosphere.


\section{Inversion of theoretical \ion{Mg}{2} {\lowercase{h \& k}} profiles}\label{Stest}

We have chosen to test the inversion code
with the \ion{Mg}{2} h and k doublet around
279~nm. This decision is not only motivated by the physical properties of this doublet,
namely, the significance of non-LTE, \gls*{prd} and scattering polarization effects in
these lines, but also because of the successful observations by the CLASP2 and CLASP2.1 missions,
whose interpretation would greatly benefit from the availability of a suitable inversion
method. To mimic the real resolution of observations, we adopted the 
spectral sampling of the IRIS instrument \citep{DePontieu2014SoPh}
for the line cores of the synthetic profiles, that is, 25.4~m\AA/pixel, whereas for the line 
wings the sampling is 8 times larger. 
In the forward calculation a thinner wavelength grid 
is generated to synthesize the profiles with sufficient accuracy.

The Stokes profiles of the \ion{Mg}{2} h and k lines can be reliably modeled
by solving the problem of the generation and transfer of polarized radiation
using a two-term Mg II atomic model (including the ground level of \ion{Mg}{2}, 
the first excited term of \ion{Mg}{2} with the two upper levels of the h and k transition, 
as well as the ground term of \ion{Mg}{3}), 
as can be easily checked by comparing the
two-term results of \cite{Tanausu2016ApJ} with the three-term results of
\cite{Tanausu2020ApJ}. Therefore, in this paper we have used a two-term atomic
model taking into account the effects of \gls*{prd}, quantum interference, and
the impact of arbitrary magnetic fields (see the cited papers for further
details on the spectral synthesis).

In this section we show the capability of the TIC to retrieve a known
model atmosphere by inverting a set of synthetic Stokes parameters. First, we apply
the inversion code directly to the Stokes profiles resulting from the synthesis, and
later, in \S.~\ref{sec:noise}, we perform the inversion after adding Gaussian noise to them. 

When modeling \gls*{rt} with scattering polarization, \gls*{prd} effects, and with
arbitrary magnetic fields, the inversion becomes very time consuming. In particular,
inverting the whole set of physical parameters at the same time becomes prohibitive.

In this paper, the full inversions are broken down into six cycles. 
In the first two cycles (hereafter, the \emph{Stokes-$I$ inversion}) 
we only fit the Stokes
$I$ profile by inverting the temperature, micro-turbulent velocity, bulk vertical
velocity ($v_{\rm ver}$), and the top boundary gas pressure 
(using 8, 3, 4, and 1 nodes, respectively).  
It has been shown \citep{delaCruz2016ApJ,delaCruz2019A&A,SainzDalda2019ApJ} that it
is possible to recover the thermal stratification of the model atmosphere 
by only fitting the intensity profile of the \ion{Mg}{2} h and k lines. We then fix these
non-magnetic model parameters and only invert the magnetic field vector in the
subsequent four cycles (hereafter, the \emph{magnetic inversion}).
This separation between Stokes-$I$ and magnetic inversions is often possible because,
for solar applications, both the magnetic field and the anisotropy are small enough
(relative to the Doppler width and the mean radiation field, respectively)
as to typically have a negligible impact on the Stokes $I$ profile
of these spectral lines. In our tests, 
the magnetic field is described by the longitudinal component $B_{\parallel}$, 
the transverse component $B_{\perp}$, and the azimuth on the plane of sky $\phi_{B}$.

The physical model of scattering polarization considered in this work is highly
multi-dimensional, and the inversion can easily get trapped in a local minimum.
In order to prevent this, we first fit the circular polarization
(usually dominated by the Zeeman effect) to retrieve $B_{\parallel}$ in the first two magnetic cycles. 
Doing it in two cycles allows us to retrieve a smooth stratification. 
As the circular polarization is only sensitive to the magnetic field in the chromosphere, 
its value at the upper and lower boundaries of the height domain cannot be well constrained 
in these cycles. 
The retrieved two-node model is a suitable and smooth initialization for the second cycle. 
Even with this strategy the third cycle, the most important in the inversion process,
can still get stuck in a local minimum. If the $\chi^2$ is not significantly reduced, this
cycle must be restarted with different initial values for $B_{\perp}$ and $\phi_{B}$. 
The last cycle then
improves the fit to the Stokes profiles by allowing more freedom in the stratification.
The four cycles of the magnetic field inversion are thus performed as follows:
\begin{itemize}
\item[1.] 2 nodes in $B_{\parallel}$ to fit only the Stokes $V$ profile. 
\item[2.] Starting from cycle 1, 4 nodes in $B_{\parallel}$ to fit only the Stokes $V$ parameter. 
\item[3.] Starting from cycle 2, 4 nodes in $B_{\parallel}$, 1 node in $B_{\perp}$, 
and 1 node in $\phi_{B}$ to fit Stokes $Q$ and $U$, and $V$, with relatively small weights
for the linear polarization.
\item[4.] Starting from cycle 3, 4 nodes in $B_{\parallel}$, 4 nodes in $B_{\perp}$, 
and 1 node in $\phi_{B}$ to fit Stokes $Q$, $U$, and $V$, with larger weights for the
linear polarization.
\end{itemize}
Following this inversion strategy, each full inversion (i.e. 6 cycles) shown in this section takes of the
order of $10^3$ CPU hours (@2.10GHz). We have tested adding one more cycle including all physical
parameters together, but the result is not significantly improved and the required computing time
is tripled.

We tested the TIC code with the \ion{Mg}{2} h and k Stokes profiles in two different magnetic
regimes: the Hanle effect around the regime of criticality, and in the ``saturation limit''.
In the former, the magnetic field produces Zeeman splittings of the order of the line's natural width.
In the latter, the magnetic field is sufficiently strong to completely relax the  
quantum coherence between magnetic sublevels, 
and the line's scattering polarization becomes only sensitive to the direction
of the magnetic field, but not to its strength.
The critical magnetic field for the onset of the Hanle effect in the \ion{Mg}{2} k line is
$\sim22$~G (remember that the h line is intrinsically non polarizable through atomic alignment due to the
angular momentum $J=1/2$ of both lower and upper levels). Therefore, the linear polarization in
the \ion{Mg}{2} k line is sensitive via the Hanle effect to magnetic field strengths
between $\sim5$ and $\sim100$~G. In \S.~\ref{sec:hanlereg}, where
we test the Hanle regime, the chosen magnetic field strength is within such limits.
In Sect.~\ref{sec:sathanlereg}, where we test the saturated Hanle regime, the chosen
magnetic field strength is larger than $100$~G.


\subsection{Hanle effect regime}\label{sec:hanlereg}

In the tests shown in this section we impose a magnetic field in the Hanle regime
for the \ion{Mg}{2} k line, that is, between $\sim5$ and $\sim100$~G.
We want to emphasize that for magnetic field strengths outside of this range (but still not
large enough to produce significant linear polarization via the Zeeman effect) the linear
polarization is completely insensitive to the magnetic field strength and the
inversion cannot possibly recover its value.

\begin{figure}[htp]
\center
\includegraphics[width=0.80\textwidth]{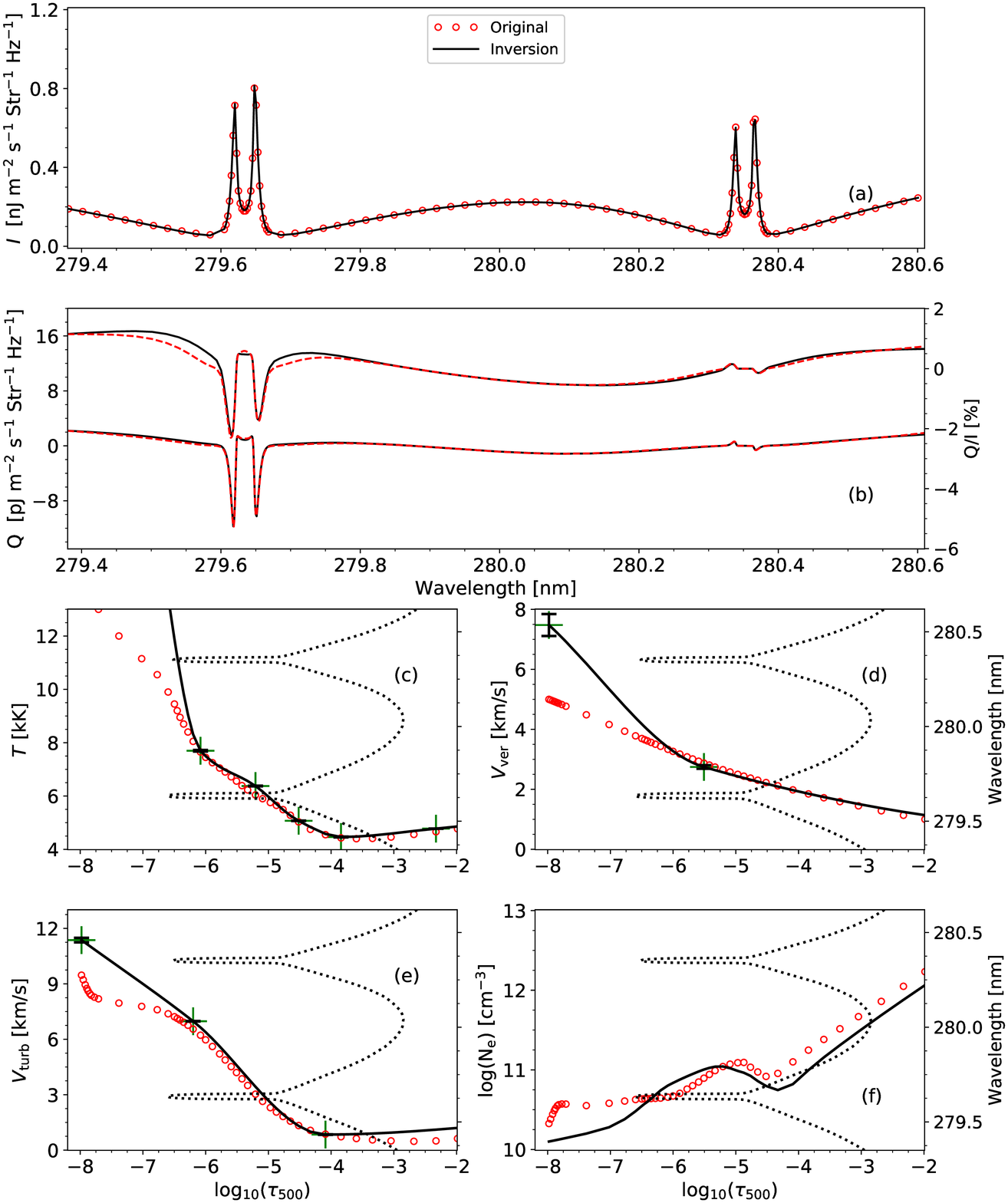}
\caption{Stokes-$I$ inversion of \ion{Mg}{2} h and k line profiles calculated in the FAL-C model
with a magnetic field in the Hanle regime.
(a) Theoretical (red circles) and fitted (solid black curve)
intensity profiles. (b) linear polarization $Q$ (bottom curves) and fractional linear polarization 
$Q/I$ (top curves) synthesized in the original unmagnetized model atmosphere 
(dashed red curves) and in the inverted atmosphere (solid black curves), for a \gls*{los}
with $\mu=0.3$. (c) Temperature,
(d) bulk vertical velocity, (e) micro-turbulent velocity, and (f) electron density
for the original (red dots) and inverted (solid black curve) atmosphere. The black
dotted curves show the optical depth (${\rm \log_{10}}(\tau_{500})$) where $\tau=1$ at
each wavelength.}
\label{fig1}
\end{figure}

\begin{figure}[htp]
\center
\includegraphics[width=0.9\textwidth]{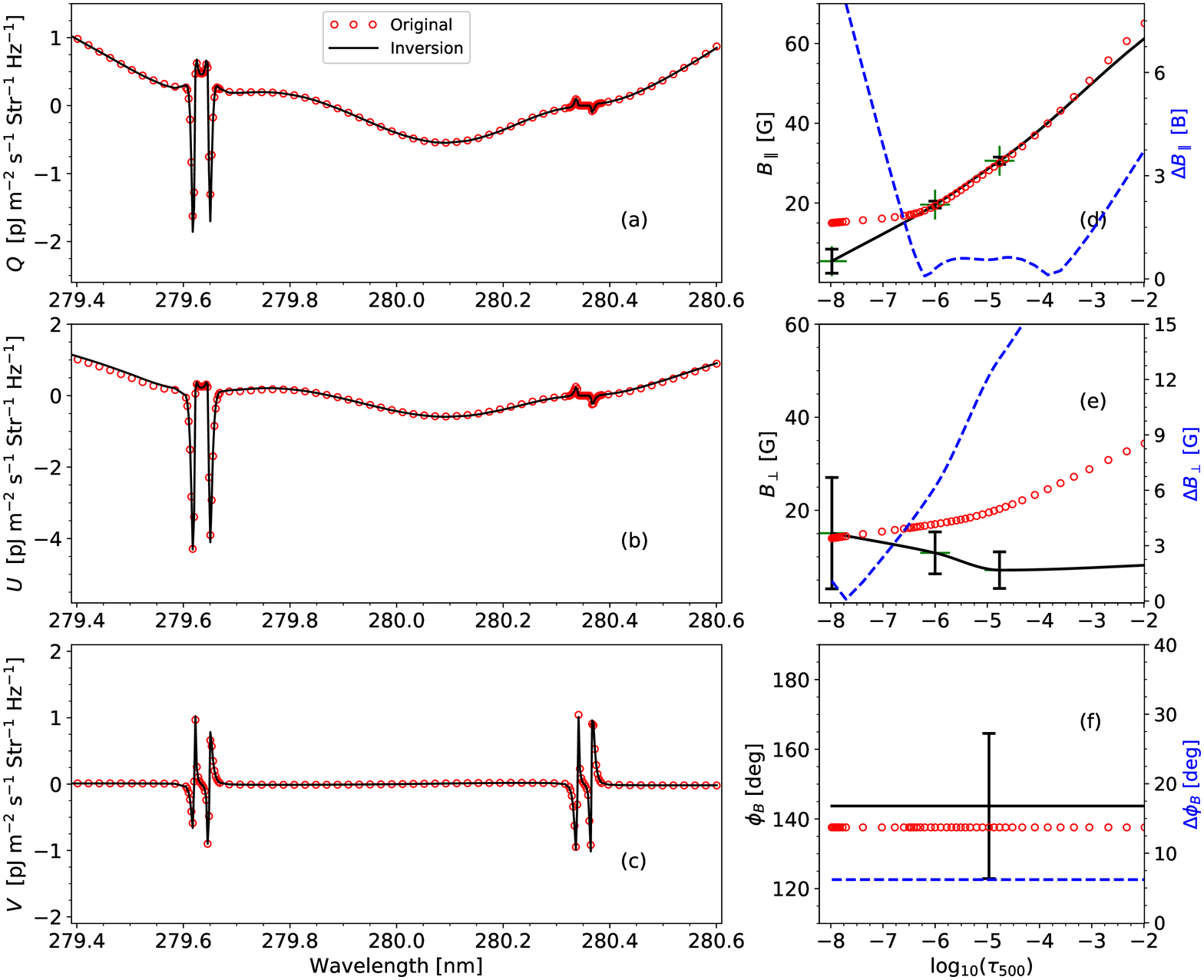}
\caption{Magnetic inversion of \ion{Mg}{2} h and k line profiles calculated in the FAL-C model 
with a magnetic field in the Hanle regime.
Linear polarization $Q$ (a) and $U$ (b), and 
circular polarization $V$ (c) profiles for the theoretical (red circles)
and fitted (black solid curves) profiles for a \gls*{los} with $\mu=0.3$.
Longitudinal magnetic field (d), 
transverse magnetic field (e), and magnetic field azimuth (f) for the original
(red circles) and inverted (black solid curves) model atmosphere.
The absolute differences between the original and inverted stratifications
are shown by the blue dashed curves and they correspond to 
the scales on the right vertical axis.}
\label{fig2}
\end{figure}

For our model atmosphere, we have taken the temperature, micro-turbulent velocity,
and electron number density of model C of \citet[][hereafter, FAL-C model,
see red circles in panels (c), (e), and (f) of Fig.~\ref{fig1}, respectively]{Fontenla1993ApJ}.
We have added a smooth stratification of the bulk vertical 
velocity (see red circles in panel (d) of Fig.~\ref{fig1}),
and an assigned stratification of the magnetic field vector
(red circles in the right column panels of Fig.~\ref{fig2}).

As explained above, we first invert the temperature, the micro-turbulent and
bulk vertical velocities, and the gas pressure at the top boundary by
fitting just the Stokes $I$ profile. The inversion is initialized with the temperature 
and micro-turbulent velocity from the model P of
\citet[][hereafter, FAL-P model]{Fontenla1993ApJ}. 
The result of this inversion is
shown in Fig.~\ref{fig1}. Panel (a) shows the theoretical
profile that represents our observation (red circles) and the fit
resulting from the inversion (black solid curve). The stratification
of the temperature (panel (c)), the bulk vertical velocity (panel (d)), the
micro-turbulent velocity (panel (e)), and the electron density (panel (f)) are
shown for the original model (red circles) and the inverted model
(black solid curve). 
The green ``$+$'' symbols in panels (c)-(e) of Fig.~\ref{fig1} 
indicate the inversion nodes
and the vertical black solid bars represent the uncertainty derived from 
the response function following Eq.~\eqref{Eq2}.
The black dotted curves in panels (c)-(f) show the height where the optical depth $\tau_\lambda$ 
for each wavelength is equal to unity 
(right vertical axis in each panel). 

This Stokes-$I$ inversion 
test was performed for a \gls*{los} with $\mu=0.3$, where $\mu$ is the
cosine of the heliocentric angle. The inversion successfully fits the theoretical
profiles and recovers the main features of the original model atmosphere above
\TAUE{-3} (note that the \ion{Mg}{2} lines are insensitive to the model
parameters at larger optical depths).

The Hanle effect is the modification of the scattering polarization at line center in the
presence of a magnetic field. Therefore, it is important to check if the inverted
atmosphere is able to reproduce the zero-field scattering polarization signals.
The red dashed curves in panel (b) of Fig.~\ref{fig1} show the Stokes $Q$ (bottom
curves) and $Q/I$ (top curves) profiles synthesized in the original model.
Likewise, the black solid curves in the same panel show the Stokes $Q$
and $Q/I$ profiles synthesized in the inverted model.
The Stokes $Q$ profile is reasonably well fitted, except for slight differences
in the center ($k_3$) and the wings ($k_1$) of the $k$ line. However, the comparison
is worse if we look at the fractional linear polarization $Q/I$ instead. Because
the intensity around the $k_1$ minimum is relatively small, the differences in
$Q/I$ are amplified in this spectral region.

In panel (c) of Fig.~\ref{fig1} we see that in the range \TAUE{-6} to $-8$ 
the inverted temperature is not exactly that of the original model.
However, the $Q$ profiles synthesized in the original and recovered models are almost the same.
Therefore, either the sensitivity to the temperature at those heights of
the atmosphere is rather small, or the temperature is to some degree
degenerate with the electron density or the micro-turbulent velocity.
Nevertheless, it is important to emphasize that the behavior
of the radiation field (and in particular its anisotropy) seems to be recovered
from the Stokes-$I$ inversion, which allows us to successively invert 
the magnetic field vector while fixing the rest of atmospheric parameters. 
We remind the reader that the results shown here correspond to an 
axially symmetric problem and thus represent an idealized case. 
In the general non-axially symmetric case, we should not expect the 
atmospheric model inverted with just the intensity to reproduce the 
polarization profiles of the actual atmosphere, even in the absence of a magnetic field.

We now perform the magnetic field inversion in four cycles as described above. 
The initial values are 0 G, 10 G, and 28.6$^\circ$ (0.5 radians) for the 
longitudinal magnetic field, the transverse magnetic field, and the azimuth, respectively.
In panels (a)-(c) of Fig.~\ref{fig2} we show the original (red circles) and 
fitted (black solid curves) polarization profiles. The panels in the right column
of Fig.~\ref{fig2} show the longitudinal component of the magnetic field (panel (d)),
the transverse component of the magnetic field (panel (e)), and the magnetic field azimuth
(panel (f)) for the original (red circles) and the inverted
(black solid curves) models, as well as the absolute difference $\Delta$ between
them (blue dashed curves and the right vertical axes).

The longitudinal component of the magnetic field is mostly
constrained by the Stokes $V$ parameter via the Zeeman effect (from \TAUA{-4.5}
and above) and the \gls*{mo} effects in the stokes $Q$ and $U$ wings (in the
range \TAUA{-3} to $\approx-5$). Between \TAUA{-7} and $\approx-4$ the absolute
error $\Delta$ on the longitudinal magnetic field component is at most 3~G. The transverse
component of the magnetic field is determined with an error smaller
than 12~G above \TAUA{-5}. The linear polarization in the k line core ($k_3$) is only
modified by the Hanle effect, while the troughs ($k_2$ and $h_2$) are modified
by both the Hanle and the \gls*{mo} effects, the latter usually being more significant.
Consequently, the transverse component
is best determined at around \TAUA{-6.5}. Finally, the inverted magnetic field azimuth
differs from the original by less than $10^\circ$.

\begin{figure}[htp]
\center
\includegraphics[width=0.9\textwidth]{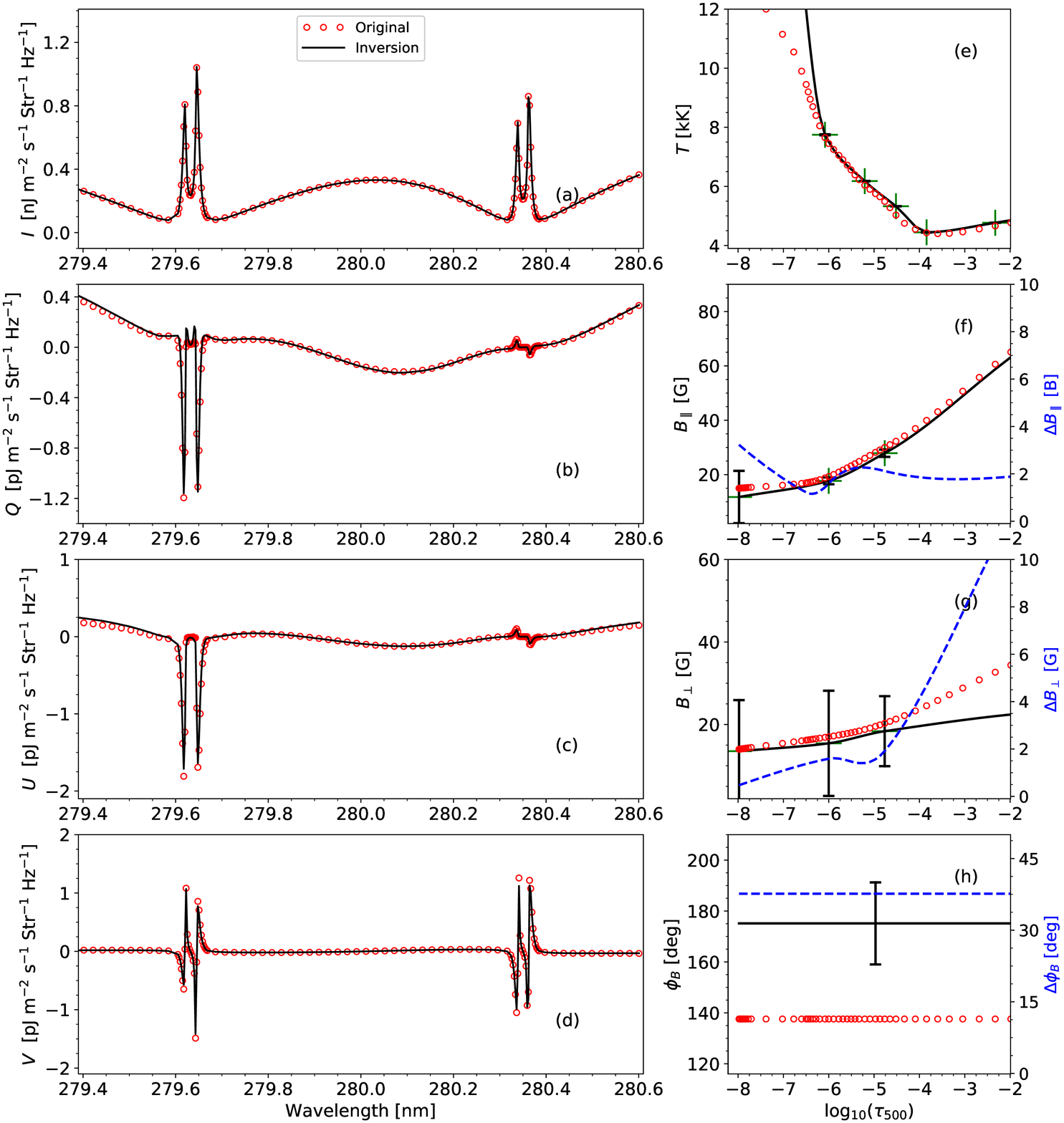}
\caption{Magnetic inversion of Stokes profiles calculated in the FAL-C model 
with a magnetic field in the Hanle regime.
Same as Fig.~\ref{fig2}, but for a \gls*{los} with $\mu=0.8$.
Here we have added panels to show the Stokes $I$ fit (a) as well as the
inversion of the temperature (e).}
\label{fig3}
\end{figure}

It is obvious that the inversion does not retrieve the original magnetic field exactly,
and it just finds the solution that best fits the Stokes profiles. Given the errors
on the temperature and density of the inverted model, it can be expected that the ensuing
changes in the radiation field and formation height of the lines will also impact the
determination of the magnetic field. This can result in errors such as those in
Fig.~\ref{fig2}. Nonetheless, the inferred magnetic field provides a
reasonably good estimation of the original magnetic field.

We repeated the same test for a \gls*{los} with $\mu=0.8$.
The results of this inversion are shown in Fig.~\ref{fig3}.
The \gls*{los} for this test is much closer to the disk center, which usually
implies a weaker linear scattering polarization. In fact,
the Stokes $Q$ and $U$ parameters are approximately a factor $4$
smaller in the line wings
and about a factor $2$ smaller at the troughs.
The weak signals of the linear polarization result in a significant decrease of 
the accuracy in the determination of the transverse magnetic field
(the inferred azimuth is $40^{\circ}$ away from the actual value).
However, the longitudinal component of the magnetic field is
constrained by the Zeeman and \gls*{mo} effects and thus is equally well determined
between \TAUA{-3} and $\approx-7$. Note that, for a \gls*{los} with $\mu=0.8$, 
the spectrum is formed deeper in the atmosphere.


\subsection{Saturated Hanle effect regime}\label{sec:sathanlereg}

In the tests shown in this section we impose a magnetic field corresponding to
the saturated Hanle regime of the \ion{Mg}{2} k line, that is, larger than
$\sim100$~G. In this regime the linear scattering polarization 
at the core of the k line is no longer
sensitive to the magnetic field strength, but only to its direction.
Therefore, in this situation, the only constraint to the magnetic field
strength comes from the Zeeman and the \gls*{mo} effects.

\begin{figure}
\center
\includegraphics[width=0.9\textwidth]{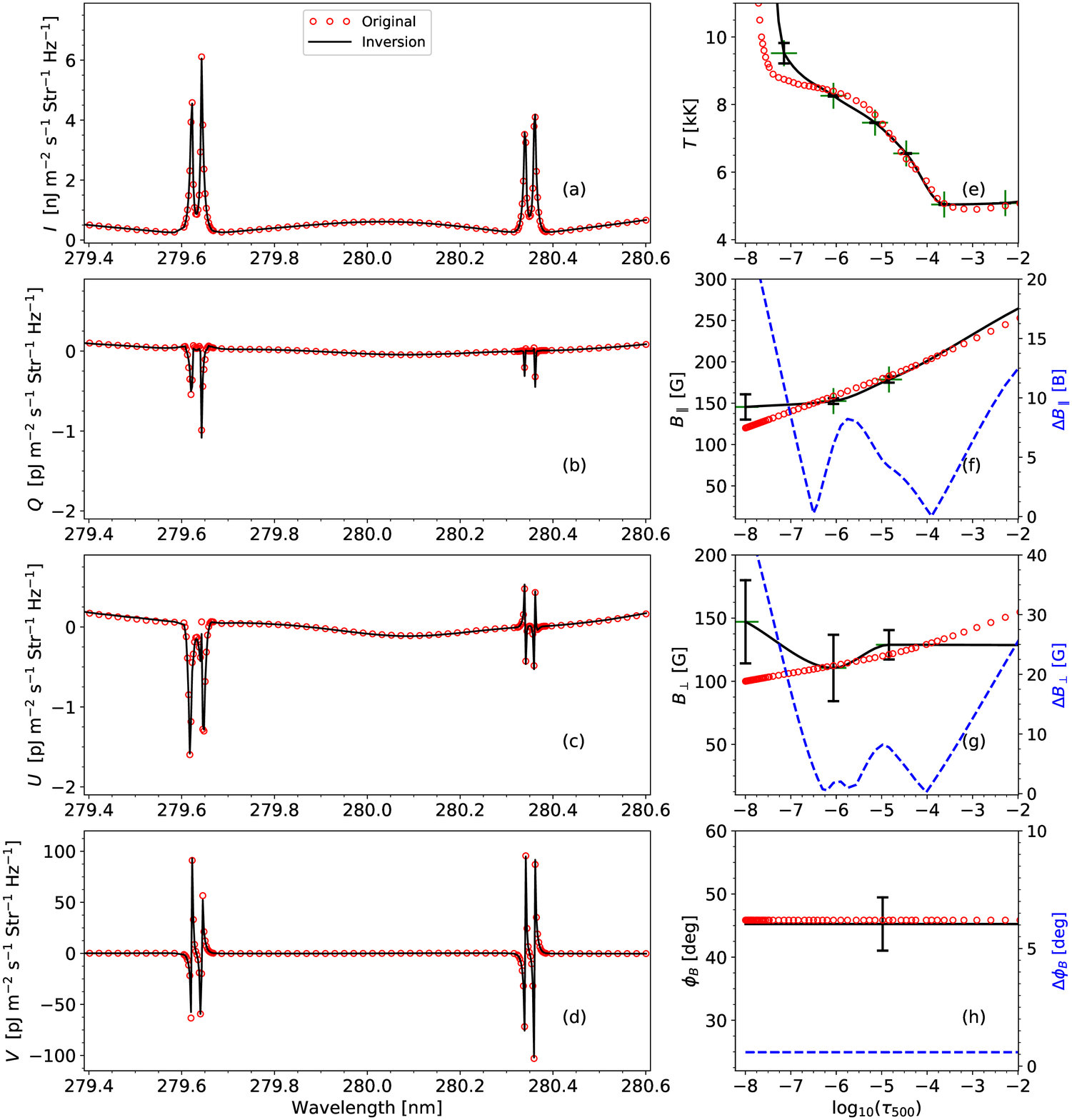}
\caption{Magnetic inversion of Stokes profiles calculated in the FAL-P model 
for a line of sight with $\mu=0.8$.
Same as Fig.~\ref{fig3}, but for the modified FAL-P model in the presence of
a stronger magnetic field in the Hanle saturation regime.}
\label{fig4}
\end{figure}

Because we are imposing a stronger magnetic field, we perform this test not
only for the FAL-C model, but also for the FAL-P model,  
as it should be more representative of plage
regions where we can find magnetic fields with such strengths. Similarly to
what we did for the FAL-C model, we added a smooth stratification of
the bulk vertical velocity, as well as a given
stratification of the magnetic field vector (red dots in panels
(e)-(h) of Fig.~\ref{fig4}). The inversion for the FAL-P model is initialized with the temperature 
and micro-turbulent velocity of the FAL-C model.

\begin{figure}
\center
\includegraphics[width=0.9\textwidth]{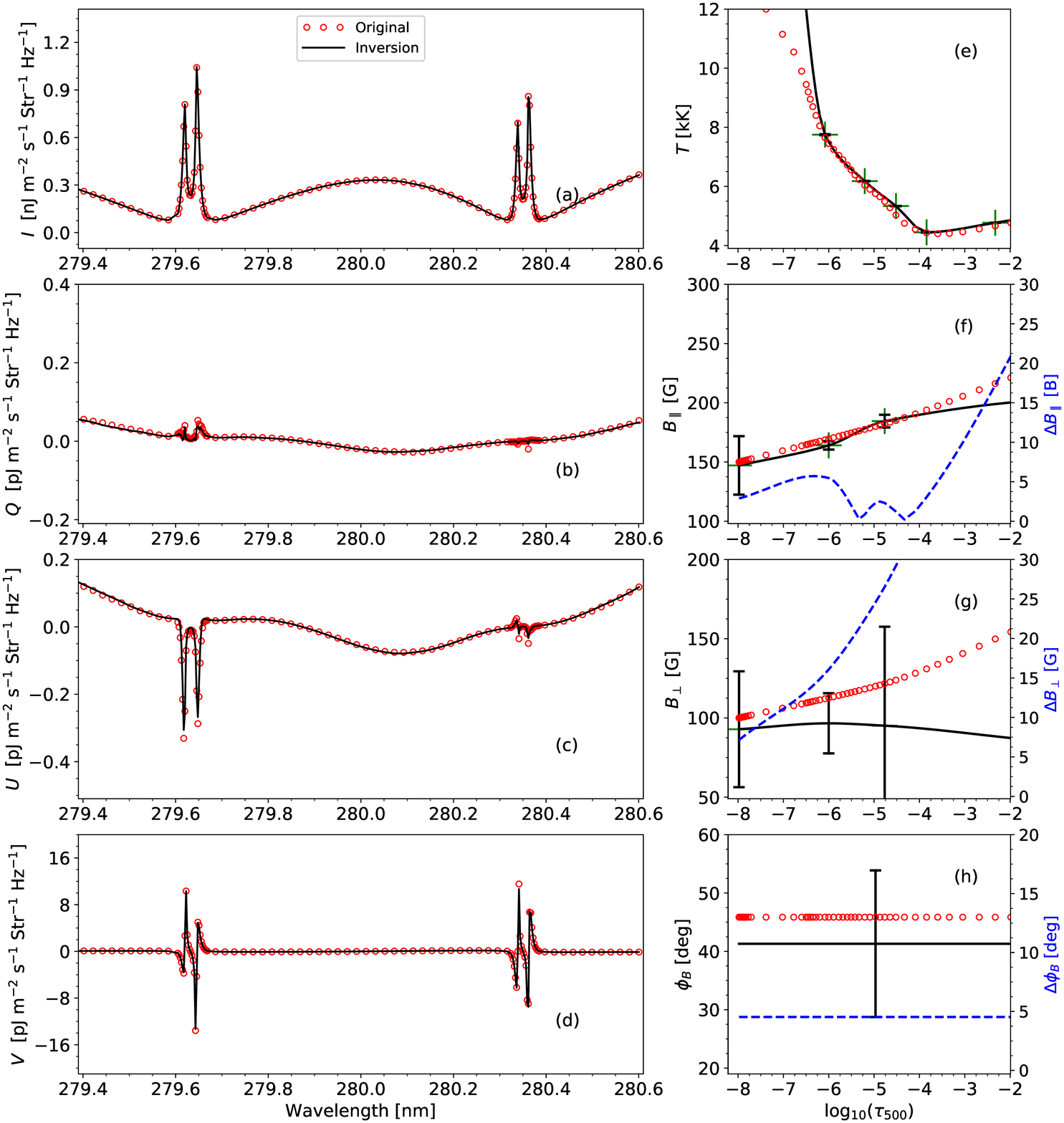}
\caption{Magnetic inversion of Stokes profiles calculated in the FAL-C model 
for a line of sight with $\mu=0.8$.
Same as Fig.~\ref{fig3}, but with a stronger magnetic field in the Hanle saturation
regime.}
\label{fig5}
\end{figure}

We follow the same inversion procedure as in \S~\ref{sec:hanlereg}, namely,
a non-magnetic inversion followed by four cycles where only the magnetic field
vector is inverted. 
We tried using the same initialization for the magnetic field as in the former test. 
However, for the FAL-C model, the initial transverse magnetic field had to be 
increased from 10 to 60~G to reach convergence in the inversion process. 
As seen in Fig. 4, the fractional linear polarization is relatively weak, which 
affects the calculation of the response function. For example, for $B_\perp=10$~G 
the difference between two syntheses for a small perturbation in the magnetic field 
is of the order of the accuracy of the forward engine.
In Figs.~\ref{fig4} and \ref{fig5} we show the inversion
results for the FAL-P and FAL-C models, respectively. In both models, the
inversion code does a good job at retrieving the longitudinal component of the magnetic
field in the region where the near wings of the \ion{Mg}{2} resonance lines are formed
(between \TAUA{-4} to $\approx-6$), constrained by both the circular polarization caused by the 
Zeeman effect and the linear polarization in the wings sensitive to the
\gls*{mo} effects. The inversion of the transverse component of the magnetic
field relies mostly on the Hanle effect in the core of the k line, and in the FAL-P
model it is relatively well determined between \TAUA{-5.0} and $\approx-6.5$. 
It is also worth mentioning that we needed to significantly
increase the weights in the cost function (Eq.~\eqref{Eq1}) for the linear polarization
in this inversion due to the relative amplitudes of the Stokes parameters. Otherwise,
the linear polarization is not well fitted and the transverse magnetic field cannot be constrained.


\subsection{Noisy Profiles}\label{sec:noise}

In the previous sections we have only considered theoretical Stokes profiles free
of noise. In this section we repeat the testof \S~\ref{sec:hanlereg}
(Hanle regime,
FAL-C model, for a \gls*{los} with $\mu=0.3$) but adding random noise
to the synthetic profiles. The added noise follows a Gaussian distribution with a
standard deviation of $\sigma = 0.2$ \SII. This corresponds to a polarimetric noise
slightly larger than $10^{-3}$ in the line wings, but better than $10^{-3}$ in the
line core. This condition is comparable to that of the
plage target of the CLASP2 mission \citep{Ishikawa2021}.

\begin{figure}
\center
\includegraphics[width=0.9\textwidth]{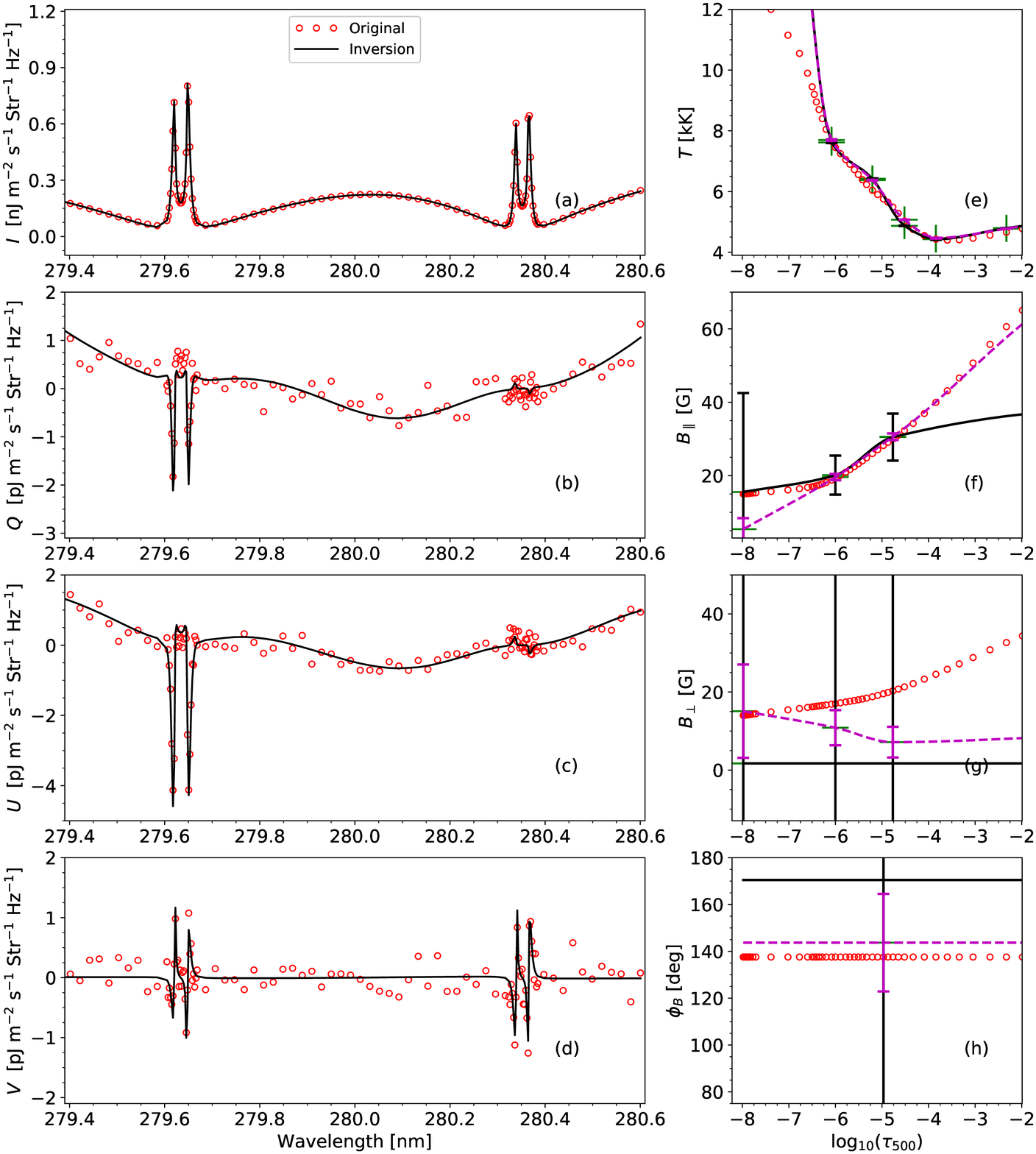}
\caption{Magnetic inversion of Stokes profiles calculated in the FAL-C model 
with a magnetic field in the Hanle regime.
Same than Fig.~\ref{fig2}, but with gaussian noise (see text) added to the
profiles. Here we have added panels to show the Stokes $I$ fit (a) as well as the
inversion of the temperature (e). The magenta curves
show the inverted model from Fig.~\ref{fig2}, that is, in the noise-free case.}
\label{fig6}
\end{figure}

With this level of noise, the wing structures in the linear polarization profiles are still
preserved. Instead, the antisymmetric linear polarization feature around the h
line core is lost in the noise. We invert these profiles with the same
strategy described above and show the results in Fig.~\ref{fig6}.
A comparison with the results from Fig.~\ref{fig2}, also shown in Fig.~\ref{fig6},
indicates that, between \TAUA{-5} and
$-6$, the inferred longitudinal magnetic field is almost
the same as in the noise-free tests. We note how, despite the poorer precision
of the inversion, leading to larger parameter uncertainties, the inferred solution
``in the mean'' is still rather accurately matching the original model.

The results of our tests are very promising in view of the possible application
of TIC to the interpretation of data from the recent the CLASP2 and CLASP2.1
missions.

 
\subsection{Longitudinal Magnetic Field}\label{sec:longB}

\begin{figure}
\center
\includegraphics[width=0.95\textwidth]{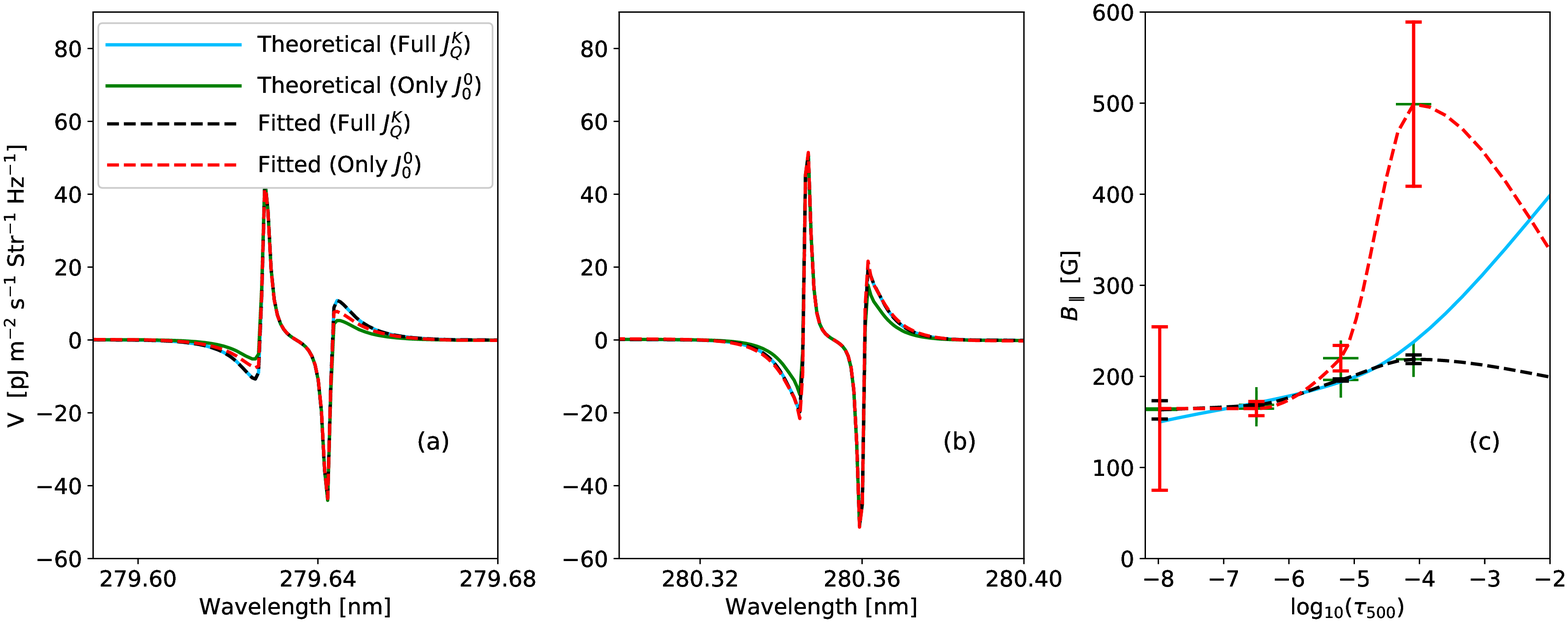}
\caption{Theoretical (solid curves) and the fit (dashed curves) Stokes $V$ profiles
of the \ion{Mg}{2} k (a) and h (b) lines. (c) input and inverted magnetic field model.
The different colors represent different cases indicated in the legend, with ``Full $J^K_Q$''
and ``Only $J^0_0$'' indicating that scattering polarization has been taking into account
and neglected, respectively.
}
\label{fig7}
\end{figure}

The circular polarization is mainly produced by the Zeeman effect and it is
only sensitive to the longitudinal component of the magnetic field. However, due to
\gls*{prd} effects, scattering polarization has a significant influence on the outer
lobes of the \ion{Mg}{2} h and k circular polarization profiles
\citep{AlsinaBallester2016ApJ,Tanausu2016ApJ}.

In this section we study the impact of neglecting atomic polarization in
the inversion of the longitudinal component of the magnetic
field. To this end, we carry out two inversions in which we only consider the
Stokes $I$ and $V$ parameters, using a FAL-C model with an imposed magnetic field for a
\gls*{los} with $\mu=$1.
For this test we fix the thermodynamic stratification and we only invert for the
longitudinal component of the magnetic field. For the first inversion, we completely neglect
scattering polarization in the synthesis of the circular polarization profile.
For the second inversion, we include all physical ingredients as for the other
tests in this paper. We find that the recovered longitudinal magnetic
field is significantly overestimated when neglecting the anisotropy of the radiation
field (see panel (c) in Fig.~\ref{fig7}). This overestimation occurs mainly below
\TAUA{-5.5}, in the region where the outer lobes of the $V$ profile are formed. 
At \TAUA{-5.2}, the difference between the original and
inverted magnetic field can reach $\sim24$~G ($\approx12$~\% relative difference).
When atomic polarization and the radiation field anisotropy are included, the
inferred magnetic field is much closer to the original stratification at
those heights where the emergent Stokes $V$ is sensitive to the magnetic field.

The black curves in the left and middle panels in Fig.~\ref{fig7} show the synthetic
profiles for the same atmospheric and magnetic field models accounting for
the Zeeman effect without atomic polarization. It is clear that, when neglecting the scattering polarization,
the outer lobes of the Stokes $V$ profile are underestimated. Therefore, the inversion
needs to increase the field strength in the lower atmosphere in order to compensate
for this effect and fit the outer lobes.


\section{Summary and discussions}\label{Scon}

In this paper we have 
presented the Tenerife Inversion Code, TIC, and tested it on
synthetic spectropolarimetric profiles to infer the magnetic and thermodynamic
structure of solar model atmospheres.
TIC is based on the HanleRT forward synthesis
engine, and takes into account all the physical mechanisms that are essential to
the modeling of strong resonance spectral lines formed in the chromosphere and transition region:
atomic polarization, magnetic field of arbitrary strength, and \gls*{prd} effects.
TIC minimizes a cost function measuring the quality of the fit to the Stokes data,
and includes a series of regularization terms to additionally constrain
the solution of this notoriously ill-posed inversion problem.

To test the inversion code, we considered a modified
version of the C and P models of \cite{Fontenla1993ApJ} to which we added a
stratified bulk vertical velocity and a magnetic 
field. With these models, we
computed the emergent
Stokes profiles of the \ion{Mg}{2} h and k lines for two different \gls*{los}.
In particular, we performed tests for two relevant regimes of the
magnetic field strength.

In the first test the magnetic field strength corresponds to the Hanle regime for the
\ion{Mg}{2} k line (the h line is intrinsically unpolarizable because
the angular momentum of its upper and lower levels is $1/2$ and thus it is
not sensitive to the Hanle effect). In this regime, the Zeeman splitting
is of the order of the natural width of the line, and the scattering polarization
is sensitive to both the strength and direction of the magnetic field. In the second
test the magnetic field strength is in the ``saturated'' Hanle regime.
Here, the Zeeman splitting is at least one order
of magnitude larger than the natural width of the line, and the scattering polarization
is only sensitive to the direction of the magnetic field. The magnetic field strength
chosen for the second test is such that the transverse Zeeman effect is still
completely negligible.

We tested the performance of the inversion with and without photon
noise and quantified how the inference of the stratification of physical parameters
is affected by the increased uncertainty.

We proposed an inversion strategy that we applied to every test shown in this paper.
This strategy follows the usual approach of first inferring only the stratification
of the thermodynamic parameters (every physical parameter but the magnetic field)
from just the intensity profile, and then invert the magnetic field with all four
Stokes profiles after fixing the rest of the physical parameters.
The inference of the transverse magnetic field component from the linear polarization profiles
can be difficult, depending on the strength and direction of the field,
and thus the inversion is iterated through several cycles in which we
incrementally converge to the final solution. We determined a good working strategy
using four magnetic inversion cycles. The convenience of performing the magnetic field
inference in this way is twofold. On the one hand, by increasing the number of
nodes and the weights for the linear polarization in incremental steps, always
using the output of the previous step as an initial condition, we approach
more stably the minimum of the cost function in Eq.~\eqref{Eq1}.
We have found that performing just one magnetic cycle
can easily lead the inversion towards a local minimum of the cost function.
On the other hand, the computing time is proportional to the number of nodes. By
finding first a rough approximation to the final solution with fewer nodes,
and then successively refine the result, we can reach the same or a better
solution in significantly less time. While the inversion strategy used
in this paper works optimally for the considered tests, it is not necessarily the
best general approach for an arbitrary set of observations. Based on the experience
acquired with these tests, it seems advisable to approach the inversion of real
spectropolarimetric data using different inversion strategies in order to find
the optimal one.

While the inversion of thermodynamic quantities inferred from just the intensity
profile is not new 
\cite[e.g.,][]{Socas-Navarro2015A&A,delaCruz2019A&A,SainzDalda2019ApJ}, 
it is an important check that every inversion code must pass before advancing to the next step of inferring
the magnetic field vector by full Stokes inversion. In all our tests
the inference of the thermodynamic quantities is achieved satisfactorily
in the range of heights corresponding to the region of
formation of the \ion{Mg}{2} h and k lines, including the extended wings
whose presence is due to the \gls*{prd} effects. The differences between
the inferred and original models in this region of formation are likely due to
model degeneracies among the parameters and to the different sensitivity of the
emergent Stokes profiles to them. Finally, it is important to observe that
assuming hydrostatic equilibrium we are not directly
inverting the stratification of the electron density,  
and the hydrogen populations are computed in LTE. 

As we mentioned above, the inference of the magnetic field vector is
significantly more complicated. In the Hanle regime (Zeeman splitting
comparable to the natural width of the line), the longitudinal magnetic field component
affects both the linear and circular spectral line polarization.
As to the linear polarization, the extended wings of the \ion{Mg}{2} h and k lines
are sensitive to the longitudinal component of the magnetic field
in the upper photosphere and low chromosphere (from \TAUA{-3} to
\TAUA{-4.5}); the circular polarization instead is sensitive to fields in
the mid-upper chromosphere (from \TAUA{-4.5} to \TAUA{-6}). Consequently, the longitudinal
component of the magnetic field is satisfactorily inferred in the
region of formation of the h and k lines of the model atmosphere.

The inference of the strength and azimuth of the transverse component of the
magnetic field is less accurate. When the contribution of the 
Zeeman effect to the linear polarization in the core of the lines 
is negligible (as in our tests) only the
linear polarization in the core of the \ion{Mg}{2} k line is sensitive to
the magnetic field vector via the Hanle effect. Because the line core forms in
the upper chromosphere (\TAUA{-7} in the model) the inference of
the stratification of the transverse field component is worse than for
the longitudinal component. Nevertheless, the inferred values
in the region of formation of the \ion{Mg}{2}
k line core are not far from the original ones. We note that
the linear polarization in the troughs of the k line is also sensitive
to the transverse component of the magnetic field via the Hanle effect,
but its sensitivity to the longitudinal magnetic field component via
the magneto-optical effects is much more significant.
The magnetic field azimuth is the parameter that has shown the strongest
dependence on the inversion strategy and, while the
inferred values are often close to the original ones, we also get
errors up to $40^\circ$ (see Fig.~\ref{fig3}) in some cases, making this
the less reliable of the inferred parameters.

We have also tested the performance of the TIC by adding photon noise
to the input data. As expected, the uncertainty associated with the
inversion (see Eq.~\eqref{Eq2}) increases and the transverse
field and its azimuth are the most affected parameters.
Even though the inverted stratification of the transverse field
is strongly impacted by the 
presence of polarimetric noise
(see Fig.~\ref{fig6}), around the formation height where the line is
sensitive to the field (between \TAUA{-5.0} and $\approx-6.5$) the
inferred magnetic strength with and without noise are reasonably close.

Finally, we presented an application where only the circular polarization
is used to infer the longitudinal component of the magnetic field. An
important conclusion is that the atomic alignment produced by the anisotropy
of the radiation field has a significant impact on the inferred field
strength, which can lead to overestimating the
actual strength of the magnetic field. 
Given that this impact of the radiation field
anisotropy to the circular polarization is consequence of the \gls*{prd}
effects \citep{Tanausu2016ApJ,AlsinaBallester2016ApJ}, we emphasize that accounting for photon
coherence effects is also crucial for the a reliable inference of the
magnetic field stratification.

In conclusion, the tests and results shown in this paper demonstrate that our
TIC code provides a viable inversion strategy to attack the complex problem
of the generation and transfer of polarization in strong resonance lines of the
chromosphere and transition region. This paves
the way toward the application of TIC to the inversion of the exciting new spectropolarimetric
observations provided by the CLASP2 missions, as well as future mission
concepts currently under development, such as the Chromospheric Magnetism
Explorer (CMEx) and the Solar Transition UV Explorer (STRUVE).
The inversion of CLASP2 data with TIC is under way, with very promising preliminary
results that will be the subject of future publications. In parallel, we will
keep improving TIC and make it available to
the community once the public version of the HanleRT forward engine is released.

\acknowledgements
We thank Andr\'{e}s Asensio Ramos and Basilio Ruiz Cobo for helpful discussions, 
and Rebecca Centeno and Han Uitenbroek for useful suggestions and comments.
We acknowledge the funding received from the European Research Council (ERC) 
under the European Union's Horizon 2020 research and innovation 
programme (ERC Advanced Grant agreement No 742265).
This material is based upon work supported by the National Center for Atmospheric Research, 
which is a major facility sponsored by the National Science Foundation under Cooperative Agreement No. 1852977.

\bibliography{TIC}{}
\bibliographystyle{aasjournal}

\end{document}